%Paper: hep-th/9303034
%From: Javier Mas <JAMAS@GAES.USC.ES>
%Date: Fri, 5 Mar 1993 11:27:08 GMT+2

%%%%%%%%%%%%%%%%%%%%%%%%%%%%%%%%%%%%%%%%%%%%%%%%%%%%%%%%%%%%%%%%
%
% This is the Plain TeX file for the paper
%
%
%    DIFFEOMORPHISMS FROM HIGHER DIMENSIONAL W-ALGEBRAS
%
%                        by
%
%   Fernando Martinez Moras Javier Mas and Eduardo Ramos
%
%
%
%%%%%%%%%%%%%%%%%%%%%%%%%%%%%%%%%%%%%%%%%%%%%%%%%%%%%%%%%%%%%%%%

%%%%%%%%%%%%%%%%%%%%%%%%%%%%%%%%%%%%%%%%%%%%%%%%%%%%%%%%%%%%%
%%%These are the macros for submission of papers to hep-th%%%
%%%The default setting is 12pt and 1 page/side but in the%%%%
%%%future it may allow people to choose also 10 pt and%%%%%%%
%%%2 pages/side.%%%%%%%%%%%%%%%%%%%%%%%%%%%%%%%%%%%%%%%%%%%%%
%%%%%%%%%%%%%%%%%%%%%%%%%%%%%%%%%%%%%%%%%%%%%%%%%%%%%%%%%%%%%
%
\def\unlockat{\catcode`\@=11}
\def\lockat{\catcode`\@=12}
\unlockat
\def\d@f@ult{} \newif\ifamsfonts \newif\ifafour
\def\m@ssage{\immediate\write16}  \m@ssage{}
\m@ssage{hep-th preprint macros.  Last modified 16/10/92 (jmf).}
\message{These macros work with AMS Fonts 2.1 (available via ftp from}
\message{e-math.ams.com).  If you have them simply hit "return"; if}
\message{you don't, type "n" now: }
\endlinechar=-1  %don't add spaces at end of line
\read-1 to\@nswer
\endlinechar=13
\ifx\@nswer\d@f@ult\amsfontstrue
    \m@ssage{(Will load AMS fonts.)}
\else\amsfontsfalse\m@ssage{(Won't load AMS fonts.)}\fi
\message{The default papersize is A4.  If you use US 8.5" x 11"}
\message{type an "a" now, else just hit "return": }
\endlinechar=-1  %don't add spaces at end of line
\read-1 to\@nswer
\endlinechar=13
\ifx\@nswer\d@f@ult\afourtrue
    \m@ssage{(Using A4 paper.)}
\else\afourfalse\m@ssage{(Using US 8.5" x 11".)}\fi
\nonstopmode
%
%%%%%%%%%%%%%%%%%%%%%%
%%%Font definitions%%%
%%%%%%%%%%%%%%%%%%%%%%
%

\font\twelverm=cmr12
\font\ninerm=cmr9
\font\sixrm=cmr6
\font\fourteenbf=cmbx12 scaled\magstep1
\font\twelvebf=cmbx12
\font\ninebf=cmbx9
\font\sixbf=cmbx6
\font\fourteeni=cmmi12 scaled\magstep1      \skewchar\fourteeni='177
\font\twelvei=cmmi12                        \skewchar\twelvei='177
\font\ninei=cmmi9                           \skewchar\ninei='177
\font\sixi=cmmi6                            \skewchar\sixi='177
\font\fourteensy=cmsy10 scaled\magstep2     \skewchar\fourteensy='60
\font\twelvesy=cmsy10 scaled\magstep1       \skewchar\twelvesy='60
\font\ninesy=cmsy9                          \skewchar\ninesy='60
\font\sixsy=cmsy6                           \skewchar\sixsy='60
\font\fourteenex=cmex10 scaled\magstep2
\font\twelveex=cmex10 scaled\magstep1

%\font\ninex=cmex9
\ifamsfonts
   
   \font\sixex=cmex7 at 6pt
   
\else
   
   \font\sixex=cmex10 at 6pt
   
\fi
\font\fourteensl=cmsl10 scaled\magstep2
\font\twelvesl=cmsl10 scaled\magstep1

\font\sevensl=cmsl10 at 7pt
\font\sixsl=cmsl10 at 6pt

\font\fourteenit=cmti12 scaled\magstep1
\font\twelveit=cmti12

\font\fourteentt=cmtt12 scaled\magstep1
\font\twelvett=cmtt12
\font\fourteencp=cmcsc10 scaled\magstep2
\font\twelvecp=cmcsc10 scaled\magstep1

\ifamsfonts
   
\else
   
\fi
\newfam\cpfam
\font\fourteenss=cmss12 scaled\magstep1
\font\twelvess=cmss12
\font\tenss=cmss10
\font\niness=cmss9

\font\sevenss=cmss8 at 7pt
\font\sixss=cmss8 at 6pt
\newfam\ssfam
\newfam\msafam \newfam\msbfam \newfam\eufam
\ifamsfonts
 \font\fourteenmsa=msam10 scaled\magstep2
 \font\twelvemsa=msam10 scaled\magstep1
 \font\tenmsa=msam10
 \font\ninemsa=msam9
 \font\sevenmsa=msam7
 \font\sixmsa=msam6
 \font\fourteenmsb=msbm10 scaled\magstep2
 \font\twelvemsb=msbm10 scaled\magstep1
 \font\tenmsb=msbm10
 \font\ninemsb=msbm9
 \font\sevenmsb=msbm7
 \font\sixmsb=msbm6
 \font\fourteeneu=eufm10 scaled\magstep2
 \font\twelveeu=eufm10 scaled\magstep1
 \font\teneu=eufm10
 \font\nineeu=eufm9
 
 \font\seveneu=eufm7
 \font\sixeu=eufm6
 \def\hexnumber@#1{\ifnum#1<10 \number#1\else
  \ifnum#1=10 A\else\ifnum#1=11 B\else\ifnum#1=12 C\else
  \ifnum#1=13 D\else\ifnum#1=14 E\else\ifnum#1=15 F\fi\fi\fi\fi\fi\fi\fi}
 \def\hexmsa{\hexnumber@\msafam}
 \def\hexmsb{\hexnumber@\msbfam} 
\fi
\newdimen\b@gheight             \b@gheight=12pt
\newcount\f@ntkey               \f@ntkey=0
\def\f@m{\afterassignment\samef@nt\f@ntkey=}
\def\samef@nt{\fam=\f@ntkey \the\textfont\f@ntkey\relax}
\def\rm{\f@m0 }
\def\mit{\f@m1 }
\def\cal{\f@m2 }
\def\it{\f@m\itfam}
\def\sl{\f@m\slfam}
\def\bf{\f@m\bffam}
\def\tt{\f@m\ttfam}
\def\caps{\f@m\cpfam}
\def\ssf{\f@m\ssfam}
\ifamsfonts
 \def\msa{\f@m\msafam}
 \def\msb{\f@m\msbfam} \let\bb=\msb
 \def\eu{\f@m\eufam}
\else
 \let \bb=\bf \let\eu=\bf
\fi
\def\fourteenpoint{\relax
    \textfont0=\fourteencp          \scriptfont0=\tenrm
      \scriptscriptfont0=\sevenrm
    \textfont1=\fourteeni           \scriptfont1=\teni
      \scriptscriptfont1=\seveni
    \textfont2=\fourteensy          \scriptfont2=\tensy
      \scriptscriptfont2=\sevensy
    \textfont3=\fourteenex          \scriptfont3=\twelveex
      \scriptscriptfont3=\tenex
    \textfont\itfam=\fourteenit     \scriptfont\itfam=\tenit
    \textfont\slfam=\fourteensl     \scriptfont\slfam=\tensl
      \scriptscriptfont\slfam=\sevensl
    \textfont\bffam=\fourteenbf     \scriptfont\bffam=\tenbf
      \scriptscriptfont\bffam=\sevenbf
    \textfont\ttfam=\fourteentt
    \textfont\cpfam=\fourteencp
    \textfont\ssfam=\fourteenss     \scriptfont\ssfam=\tenss
      \scriptscriptfont\ssfam=\sevenss
    \ifamsfonts
       \textfont\msafam=\fourteenmsa     \scriptfont\msafam=\tenmsa
         \scriptscriptfont\msafam=\sevenmsa
       \textfont\msbfam=\fourteenmsb     \scriptfont\msbfam=\tenmsb
         \scriptscriptfont\msbfam=\sevenmsb
       \textfont\eufam=\fourteeneu     \scriptfont\eufam=\teneu
         \scriptscriptfont\eufam=\seveneu \fi
    \samef@nt
    \b@gheight=14pt
    \setbox\strutbox=\hbox{\vrule height 0.85\b@gheight
                                depth 0.35\b@gheight width\z@ }}
\def\twelvepoint{\relax
    \textfont0=\twelverm          \scriptfont0=\ninerm
      \scriptscriptfont0=\sixrm
    \textfont1=\twelvei           \scriptfont1=\ninei
      \scriptscriptfont1=\sixi
    \textfont2=\twelvesy           \scriptfont2=\ninesy
      \scriptscriptfont2=\sixsy
    \textfont3=\twelveex          %\scriptfont3=\ninex
      \scriptscriptfont3=\sixex
    \textfont\itfam=\twelveit    %\scriptfont\itfam=\nineit
    \textfont\slfam=\twelvesl    %\scriptfont\slfam=\ninesl
      \scriptscriptfont\slfam=\sixsl
    \textfont\bffam=\twelvebf     \scriptfont\bffam=\ninebf
      \scriptscriptfont\bffam=\sixbf
    \textfont\ttfam=\twelvett
    \textfont\cpfam=\twelvecp
    \textfont\ssfam=\twelvess     \scriptfont\ssfam=\niness
      \scriptscriptfont\ssfam=\sixss
    \ifamsfonts
       \textfont\msafam=\twelvemsa     \scriptfont\msafam=\ninemsa
         \scriptscriptfont\msafam=\sixmsa
       \textfont\msbfam=\twelvemsb     \scriptfont\msbfam=\ninemsb
         \scriptscriptfont\msbfam=\sixmsb
       \textfont\eufam=\twelveeu     \scriptfont\eufam=\nineeu
         \scriptscriptfont\eufam=\sixeu \fi
    \samef@nt
    \b@gheight=12pt
    \setbox\strutbox=\hbox{\vrule height 0.85\b@gheight
                                depth 0.35\b@gheight width\z@ }}
\twelvepoint
%
%%%%%%%%%%%%%%%%%
%%%Basic skips%%%
%%%%%%%%%%%%%%%%%
%
\baselineskip = 15pt plus 0.2pt minus 0.1pt %was 20pt ...
\lineskip = 1.5pt plus 0.1pt minus 0.1pt
\lineskiplimit = 1.5pt
\parskip = 6pt plus 2pt minus 1pt
\interlinepenalty=50
\interfootnotelinepenalty=5000
\predisplaypenalty=9000
\postdisplaypenalty=500
\hfuzz=1pt
\vfuzz=0.2pt
\dimen\footins=24 truecm % 8 truein in SB
\ifafour
 \hsize=16cm \vsize=22cm
\else
 \hsize=6.5in \vsize=9in
\fi
%
%%%%%%%%%%%%%%%
%%%Footnotes%%%
%%%%%%%%%%%%%%%
%
\skip\footins=\medskipamount
\newcount\fnotenumber
\def\clearfnotenumber{\fnotenumber=0} \clearfnotenumber
\def\fnote{\global\advance\fnotenumber by1 \generatefootsymbol
 \footnote{$^{\footsymbol}$}}
\def\fd@f#1 {\xdef\footsymbol{\mathchar"#1 }}
\def\generatefootsymbol{\iffrontpage\ifcase\fnotenumber
\or \fd@f 279 \or \fd@f 27A \or \fd@f 278 \or \fd@f 27B
\else  \fd@f 13F \fi
\else\xdef\footsymbol{\the\fnotenumber}\fi}
%
%%%%%%%%%%%%%%%%%%%%%%%%%%%%%
%%%Sections and Appendices%%%
%%%%%%%%%%%%%%%%%%%%%%%%%%%%%
%
\newcount\secnumber \newcount\appnumber
\def\clearappnumber{\appnumber=64} \def\clearsecnumber{\secnumber=0}
\clearsecnumber \clearappnumber
\newif\ifs@c % this is true if within a section as opposed to an appendix
\newif\ifs@cd % this is true if the article is being section'd
\s@cdtrue % this is the default
\def\unsectioned{\s@cdfalse\let\section=\subsection}
\newskip\sectionskip         \sectionskip=\medskipamount
\newskip\headskip            \headskip=8pt plus 3pt minus 3pt
\newdimen\sectionminspace    \sectionminspace=10pc
\def\Titlestyle#1{\par\begingroup \interlinepenalty=9999
     \leftskip=0.02\hsize plus 0.23\hsize minus 0.02\hsize
     \rightskip=\leftskip \parfillskip=0pt
     \advance\baselineskip by 0.5\baselineskip%this is a test...
     \hyphenpenalty=9000 \exhyphenpenalty=9000
     \tolerance=9999 \pretolerance=9000
     \spaceskip=0.333em \xspaceskip=0.5em
     \fourteenpoint
  \noindent #1\par\endgroup }
\def\titlestyle#1{\par\begingroup \interlinepenalty=9999
     \leftskip=0.02\hsize plus 0.23\hsize minus 0.02\hsize
     \rightskip=\leftskip \parfillskip=0pt
     \hyphenpenalty=9000 \exhyphenpenalty=9000
     \tolerance=9999 \pretolerance=9000
     \spaceskip=0.333em \xspaceskip=0.5em
     \fourteenpoint
   \noindent #1\par\endgroup }
\def\spacecheck#1{\dimen@=\pagegoal\advance\dimen@ by -\pagetotal
   \ifdim\dimen@<#1 \ifdim\dimen@>0pt \vfil\break \fi\fi}
\def\section#1{\cleareqnumber \s@ctrue \global\advance\secnumber by1
   \par \ifnum\the\lastpenalty=30000\else
   \penalty-200\vskip\sectionskip \spacecheck\sectionminspace\fi
   \noindent {\caps\enspace\S\the\secnumber\quad #1}\par
   \nobreak\vskip\headskip \penalty 30000 }
\def\undertext#1{\vtop{\hbox{#1}\kern 1pt \hrule}}
\def\subsection#1{\par
   \ifnum\the\lastpenalty=30000\else \penalty-100\smallskip
   \spacecheck\sectionminspace\fi
   \noindent\undertext{#1}\enspace \vadjust{\penalty5000}}

\def\appendix#1{\cleareqnumber \s@cfalse \global\advance\appnumber by1
   \par \ifnum\the\lastpenalty=30000\else
   \penalty-200\vskip\sectionskip \spacecheck\sectionminspace\fi
   \noindent {\caps\enspace Appendix \char\the\appnumber\quad #1}\par
   \nobreak\vskip\headskip \penalty 30000 }

\def\refs{\begingroup \par\penalty-100\medskip \spacecheck\sectionminspace
   \line{\fourteencp\hfil REFERENCES\hfil}%
\nobreak\vskip\headskip \frenchspacing }
\def\endrefs{\par\endgroup}
%--- Note added
%
%%%%%%%%%%%%%%%%%%%%%%%%%%%%%%%%%
%%%Running heads and footlines%%%
%%%%%%%%%%%%%%%%%%%%%%%%%%%%%%%%%
%
\newif\iffrontpage \frontpagefalse
\headline={\hfil}
\footline={\iffrontpage\hfil\else \hss\twelverm
-- \folio\ --\hss \fi }
%
%%%%%%%%%%%%%%%%
%%%Title page%%%
%%%%%%%%%%%%%%%%
%
\newskip\frontpageskip \frontpageskip=12pt plus .5fil minus 2pt
\def\titlepage{\global\frontpagetrue\hrule height\z@ \relax
               \pubblock\relax }
\def\endtitlepage{\vfil\break\clearfnotenumber\frontpagefalse}
\def\title#1{\vskip\frontpageskip\Titlestyle{\caps #1}\vskip3\headskip}
\def\author#1{\vskip.5\frontpageskip\titlestyle{\caps #1}\nobreak}
\def\and{\par\kern 5pt \centerline{\sl and}}

\def\authors{\vskip\frontpageskip\noindent}
\def\address#1{\par\kern 5pt\titlestyle{\it #1}}
\def\andaddress{\par\kern 5pt \centerline{\sl and} \address}
\def\addresses{\vskip\frontpageskip\noindent\interlinepenalty=9999}
\def\abstract#1{\par\dimen@=\prevdepth \hrule height\z@ \prevdepth=\dimen@
   \vskip\frontpageskip\spacecheck\sectionminspace
   \centerline{\fourteencp ABSTRACT}\vskip\headskip
   {\noindent #1}}

\def\email#1{\fnote{\tentt e-mail: #1}}

%
%%%%%%%%%%%%%%%%%%%%
%%%some addresses%%%
%%%%%%%%%%%%%%%%%%%%
%

%

%

%

%
%%%%%%%%%%%%%%%%
%%%References%%%
%%%%%%%%%%%%%%%%
%
\newcount\refnumber \def\clearrefnumber{\refnumber=0}  \clearrefnumber
\newwrite\R@fs                              %This opens a file .refs with
\immediate\openout\R@fs=\jobname.refs %the references in order of
                                            %appearance.
\def\closerefs{\immediate\closeout\R@fs} %close file so that TeX can read it
\def\refsout{\closerefs\refs
\unlockat
\input\jobname.refs
\lockat
\endrefs}
\def\refitem#1{\item{{\bf #1}}}%just bolds it so that \bf does not expand
\def\ifundefined#1{\expandafter\ifx\csname#1\endcsname\relax}
\def\[#1]{\ifundefined{#1R@FNO}%
\global\advance\refnumber by1%
\expandafter\xdef\csname#1R@FNO\endcsname{[\the\refnumber]}%
\immediate\write\R@fs{\noexpand\refitem{\csname#1R@FNO\endcsname}%
\noexpand\csname#1R@F\endcsname}\fi{\bf \csname#1R@FNO\endcsname}}
\def\refdef[#1]#2{\expandafter\gdef\csname#1R@F\endcsname{{#2}}}
%
%%%%%%%%%%%%%%%
%%%Equations%%%
%%%%%%%%%%%%%%%
%
\newcount\eqnumber \def\cleareqnumber{\eqnumber=0}
\newif\ifal@gn \al@gnfalse  % this is true if within an \eqalignno
\def\veqnalign#1{\al@gntrue \vbox{\eqalignno{#1}} \al@gnfalse}
\def\eqnalign#1{\al@gntrue \eqalignno{#1} \al@gnfalse}
\def\(#1){\relax%
\ifundefined{#1@Q}
 \global\advance\eqnumber by1
 \ifs@cd
  \ifs@c
   \expandafter\xdef\csname#1@Q\endcsname{{%
\noexpand\rm(\the\secnumber .\the\eqnumber)}}
  \else
   \expandafter\xdef\csname#1@Q\endcsname{{%
\noexpand\rm(\char\the\appnumber .\the\eqnumber)}}
  \fi
 \else
  \expandafter\xdef\csname#1@Q\endcsname{{\noexpand\rm(\the\eqnumber)}}
 \fi
 \ifal@gn
    & \csname#1@Q\endcsname
 \else
    \eqno \csname#1@Q\endcsname
 \fi
\else%
\csname#1@Q\endcsname\fi\global\let\@Q=\relax}
%
%%%%%%%%%%%%%%%%%
%%%Mathematica%%%
%%%%%%%%%%%%%%%%%
%
\newif\ifm@thstyle \m@thstylefalse
\def\mathstyle{\m@thstyletrue}
\def\proclaim#1#2\par{\smallbreak\begingroup%        small --> med???
\advance\baselineskip by -0.25\baselineskip%
\advance\belowdisplayskip by -0.35\belowdisplayskip%
\advance\abovedisplayskip by -0.35\abovedisplayskip%
    \noindent{\caps#1.\enspace}{#2}\par\endgroup%
\smallbreak}%--- defs, thms, ...                     small --> med???
\def\m@kem@th<#1>#2#3{%
\ifm@thstyle \global\advance\eqnumber by1
 \ifs@cd
  \ifs@c
   \expandafter\xdef\csname#1\endcsname{{%
\noexpand #2\ \the\secnumber .\the\eqnumber}}
  \else
   \expandafter\xdef\csname#1\endcsname{{%
\noexpand #2\ \char\the\appnumber .\the\eqnumber}}
  \fi
 \else
  \expandafter\xdef\csname#1\endcsname{{\noexpand #2\ \the\eqnumber}}
 \fi
 \proclaim{\csname#1\endcsname}{#3}
\else
 \proclaim{#2}{#3}
\fi}
\def\Thm<#1>#2{\m@kem@th<#1M@TH>{Theorem}{\sl#2}}%--- Theorem
\def\Prop<#1>#2{\m@kem@th<#1M@TH>{Proposition}{\sl#2}}%--- Proposition
\def\Def<#1>#2{\m@kem@th<#1M@TH>{Definition}{\rm#2}}%--- Definition
\def\Lem<#1>#2{\m@kem@th<#1M@TH>{Lemma}{\sl#2}}%--- Lemma
\def\Cor<#1>#2{\m@kem@th<#1M@TH>{Corollary}{\sl#2}}%--- Corollary
\def\Conj<#1>#2{\m@kem@th<#1M@TH>{Conjecture}{\sl#2}}%--- Conjecture
\def\Rmk<#1>#2{\m@kem@th<#1M@TH>{Remark}{\rm#2}}%--- Remark
\def\Exm<#1>#2{\m@kem@th<#1M@TH>{Example}{\rm#2}}%--- Example
\def\Qry<#1>#2{\m@kem@th<#1M@TH>{Query}{\it#2}}%--- Query
%
%--- Proof
%

%
\def\<#1>{\csname#1M@TH\endcsname}
%
%%%%%%%%%%%%%%%%%%%
%%%Abbreviations%%%
%%%%%%%%%%%%%%%%%%%
%
\def\ref#1{{\bf [#1]}}%--- [ref]
%--- et al.
\def\ie{{\it i.e.\/}}%--- i.e.
%--- e.g.
%--- Cf.
%--- cf.
 %--- double left quote
%--- th as in fifth
\def\nl{\hfil\break}%--- new line
%--- 1/2
%
%%%%%%%%%%%%%%%%%
%%%Mathematics%%%
%%%%%%%%%%%%%%%%%
%
%--- def over =
%--- Halmos Q.E.D.

%--- implies
%--- is implied by
%--- if and only if
\def\lapprox{\hbox{\lower3pt\hbox{$\buildrel<\over\sim$}}}% approx lt
\def\gapprox{\hbox{\lower3pt\hbox{$\buildrel<\over\sim$}}}% approx gt
\def\quotient#1#2{#1/\lower0pt\hbox{${#2}$}}%--- factor objects
\ifamsfonts
 \mathchardef\empty="0\hexmsb3F %--- better empty set than \emptyset
 \mathchardef\lsemidir="2\hexmsb6E % semidirect |x
 \mathchardef\rsemidir="2\hexmsb6F % semidirect x|
\else
 \let\empty=\emptyset
 \def\lsemidir{\mathbin{\hbox{\hskip2pt\vrule height 5.7pt depth -.3pt
    width .25pt\hskip-2pt$\times$}}}
 \def\rsemidir{\mathbin{\hbox{$\times$\hskip-2pt\vrule height 5.7pt
    depth -.3pt width .25pt\hskip2pt}}}
\fi
%
%--- injective map
%--- surjective map
%--- bijective map
\def\to{\rightarrow}%--- mapping
%--- long mapping
%--- isom over -->
%--- just an abbrev.
%

%
 %--- commutative diagram macro
 %--- map in complex
%
 %--- reals
 %--- complex nos.
 %--- quaternions
\def\integ{\mathord{\bb Z}} %--- integers
 %--- rationals
 %--- naturals
 %--- ground field
%
%--- Hom(omorphisms)
%--- tr(ace)
\def\Tr{\mathop{\rm Tr}}%--- Tr(ace)
%--- End(omorphisms)
%--- Mor(phisms)
%--- Aut(omorphisms)
%--- aut(omorphisms)
%--- supertrace
%--- superdeterminant
%--- kernel
%--- cokernel
%--- image
%
\def\underrightarrow#1{\vtop{\ialign{##\crcr
      $\hfil\displaystyle{#1}\hfil$\crcr
      \noalign{\kern-\p@\nointerlineskip}
      \rightarrowfill\crcr}}} %--- modification of \overrightarrow
\def\underleftarrow#1{\vtop{\ialign{##\crcr
      $\hfil\displaystyle{#1}\hfil$\crcr
      \noalign{\kern-\p@\nointerlineskip}
      \leftarrowfill\crcr}}}  %--- modification of \overleftarrow

\def\comm#1#2{\left[#1\, ,\,#2\right]}%--- [ , ]
%--- { , }
%--- [ , }
%
\def\lied#1#2{{\cal L}_{#1}{#2}}%--- Lie derivative
%--- vartnl derivative
\def\pder#1#2{{{\partial #1}\over{\partial #2}}}%--- partial derivative
%--- full derivative
%
%%%%%%%%%%%%%%
%%%Journals%%%
%%%%%%%%%%%%%%
%

\def\NPB#1#2#3{{\sl Nucl. Phys.} {\bf B#1} (#2) #3}

\def\CMP#1#2#3{{\sl Comm. Math. Phys.} {\bf #1} (#2) #3}

\def\PLA#1#2#3{{\sl Phys. Lett.} {\bf #1A} (#2) #3}
\def\PLB#1#2#3{{\sl Phys. Lett.} {\bf #1B} (#2) #3}
\def\JMP#1#2#3{{\sl J. Math. Phys.} {\bf #1} (#2) #3}

\def\AoP#1#2#3{{\sl Ann. of Phys.} {\bf #1} (#2) #3}

\def\FAP#1#2#3{{\sl Funkt. Anal. Prilozheniya} {\bf #1} (#2) #3}
\def\FAaIA#1#2#3{{\sl Functional Analysis and Its Application} {\bf #1} (#2)
#3}

\def\Invm#1#2#3{{\sl Invent. math.} {\bf #1} (#2) #3}

\def\IJMPA#1#2#3{{\sl Int. J. Mod. Phys.} {\bf A#1} (#2) #3}
\def\AdM#1#2#3{{\sl Advances in Math.} {\bf #1} (#2) #3}

\def\TMP#1#2#3{{\sl Theor. Mat. Phys.} {\bf #1} (#2) #3}

\def\JSM#1#2#3{{\sl J. Soviet Math.} {\bf #1} (#2) #3}

\def\PJAS#1#2#3{{\sl Proc. Jpn. Acad. Sci.} {\bf #1} (#2) #3}
\def\JPSJ#1#2#3{{\sl J. Phys. Soc. Jpn.} {\bf #1} (#2) #3}
\def\JETPL#1#2#3{{\sl  Sov. Phys. JETP Lett.} {\bf #1} (#2) #3}

\lockat

%%%%%%%%%%%%%%%%%%%%%%%%%%%%%%%%%%%%%%%%%%%%%%%%%%%%%%%%%%%%%%%%
\mathstyle
\overfullrule=0pt
\unsectioned
%   These are the local macros
%
\def\d{\partial}
\let\pb=\anticomm

\def\fr#1/#2{\hbox{${#1}\over{#2}$}}
\def\Fr#1/#2{{{#1}\over{#2}}}

\def\Tr{{\rm Tr\,}}

 %--- dual pairing

%--- regular terms in OPE
\def\ope#1#2{{{#2}\over{\ifnum#1=1 {x-y} \else {(x-y)^{#1}}\fi}}}%--- OPEs
%--- mth full
					       %derivative
%--- mth
							      %partial
							      %derivative

\def\om{\ominus}
\def\GD{{\cal GD}}
\def\Sch{{\cal S}}
\def\W{{\ssf W}}
\def\w{{\ssf w}}

\refdef[Adler]{M.~Adler, \Invm{50}{1981}{403}.}
\refdef[MaRa]{Yu.~I.~Manin and A.~O.~Radul, \CMP{98}{1985}{65}.}
\refdef[DickeyI]{L.~A.~Dickey, {\sl Lectures in field theoretical
Lagrange-Hamiltonian formalism}, (unpublished).}
\refdef[DickeyII]{L.~A.~Dickey, \CMP{87}{1982}{127}.}
\refdef[DickeyIII]{L.~A.~Dickey, {\sl Integrable equations and Hamiltonian
systems}, World Scientific Publ.~Co.}
\refdef[GD]{I.~M.~Gel'fand and L.~A.~Dickey, {\sl A family of Hamiltonian
structures connected with integrable nonlinear differential equations},
Preprint 136, IPM AN SSSR, Moscow (1978).}
\refdef[DS]{V.~G.~Drinfel'd and V.~V.~Sokolov, \JSM{30}{1984}{1975}.}
\refdef[KW]{B.~A.~Kupershmidt and G.~Wilson, \Invm{62}{1981}{403}.}
\refdef[Mag]{F.~Magri, \JMP{19}{1978}{1156}.}
\refdef[Dickeypc]{L.~A.~Dickey, private communication.}
\refdef[STS]{M.~A.~Semenov-Tyan-Shanski\u\i, \FAaIA{17}{1983}{259}.}
\refdef[LM]{D.~R.~Lebedev and Yu.~I.~Manin, \FAP{13}{1979}{40}.}
\refdef[Uniw]{J.~M.~Figueroa-O'Farrill and E.~Ramos, {\sl Existence
and Uniqueness of the Universal $W$-Algebra}, to appear in
the {\sl J.~Math.~Phys.}}
\refdef[FL]{V.~A.~Fateev and S.~L.~Lykyanov, \IJMPA{3}{1988}{507}.}
\refdef[DFIZ]{P.~Di Francesco, C.~Itzykson, and J.-B.~Zuber,
\CMP{140}{1991}{543}.}
\refdef[KR]{V.~G.~Ka{\v c} and A.~C.~Raina, {\sl Lectures on highest
weight representations of infinite dimensional Lie algebras}, World
Scientific, etc...}
\refdef[KP]{E.~Date, M.~Jimbo, M.~Kashiwara, and T.~Miwa
\PJAS{57A}{1981}{387}; \JPSJ{50}{1981}{3866}.}
\refdef[DickeyKP]{L.~A.~Dickey, {\sl Annals of the New York Academy of
Science} {\bf 491} (1987) 131.}
\refdef[WKP]{J.~M.~Figueroa-O'Farrill, J.~Mas, and E.~Ramos,
\PLB{266}{1991}{298}.}
\refdef[Yama]{K. Yamagishi, \PLB{259}{1991}{436}.}
\refdef[YuWu]{F. Yu and Y.-S. Wu, {\sl Hamiltonian Structure,
(Anti-)Self-Adjoint Flows in KP Hierarchy and the $W_{1+\infty}$ and
$W_\infty$ Algebras}, Utah Preprint, January 1991.}
\refdef[ClassLim]{J. M. Figueroa-O'Farrill and E. Ramos, {\sl The
Classical Limit of $W$-Algebras}, \PLB{282}{1992}{357},({\tt hep-th/9202040}).}
\refdef[KoSt]{B. Kostant and S. Sternberg, \AoP{176}{1987}{49}.}
\refdef[TakaTake]{K. Takasaki and T. Takebe, {\sl SDIFF(2) KP
Hierarchy}, Preprint RIMS-814, December 1991.}
\refdef[KoGi]{Y. Kodama, \PLA{129}{1988}{223},
\PLA{147}{1990}{477};\nl
Y. Kodama and J. Gibbons, \PLA{135}{1989}{167}.}
\refdef[Bakas]{I. Bakas, \CMP{134}{1990}{487}.}
\refdef[Pope]{C. N. Pope, L. J. Romans, and X. Shen,
\PLB{242}{1990}{401}.}
\refdef[Lang]{S. Lang, {\sl Algebra}, Second Edition, Addison-Wesley
1984.}
\refdef[Radul]{A. O. Radul, \JETPL{50}{1989}{373}.}
\refdef[Morozov]{A. Morozov, \NPB{357}{1991}{619}.}
\refdef[LPSW]{H. Lu, C. N. Pope, X. Shen, and X. J. Wang, {\sl The
complete structure of $W_N$ from $W_\infty$ at $c=-2$}, Texas A\& M
Preprint CPT TAMU-33/91 (May 1991).}
\refdef[Winfty]{C. N. Pope, L. J. Romans, and X. Shen,
\PLB{236}{1990}{173},\PLB{242}{1990}{401},\NPB{339}{1990}{191};\nl
I. Bakas, \CMP{134}{1990}{487}.}
\refdef[Zam]{A. B. Zamolodchikov, \TMP{65}{1986}{1205}.}
\refdef[Douglas]{M. R. Douglas, \PLB{238}{1990}{176}.}
\refdef[Guill]{V. W. Guillemin, \AdM{10}{1985}{131}.}
\refdef[Wod]{M. Wodzicki,{\sl Noncommutative Residue} in {\sl K-Theory,
Arithmetic and Geometry}. Ed. Yu. I. Manin. Lectures Notes in
Mathematics 1289, Springer-Verlag.}
\refdef[Shubin]{M. A. Shubin, {\sl Pseudodifferential operators
and spectral Theory}, Springer-Verlag.}
\refdef[Class]{J. M. Figueroa-O'Farrill and E. Ramos, {\sl
Classical $W$-algebras from dispersionless Lax hierarchies
}, Preprint-KUL-TF-92/6, June 1992.}
\refdef[Mamoramos]{F.Mart{\'\i}nez-Mor{\'a}s and E. Ramos,
{\sl Higher Dimensional Classical W-algebras }, Preprint-KUL-TF-92/19 and
US-FT/6-92,({\tt hep-th9206040}).}
\refdef[Hull2]{C.M. Hull, {\sl W-geometry}, Preprint-QMW-92-6,
({\tt hep-th/9211113}).}
\refdef[Nijenhuis]{K. H. Bhaskara and K. Viswanath,
{\sl Poisson Algebras and Poisson Manifolds}, Pitman Research Notes
in Math. Series, vol. 174, Longman.}
\refdef[Fradkin]{E. S. Fradkin and M. A. Vasil'ev, {\sl Ann. of
Phys. 177} (1987) 199.}
\refdef[FradBerends] {E.S. Fradkin and M.A. Vasil'ev, \AoP{177}{1987}{199} ;\nl
F.A. Berends, G.J.H. Burgers and H. van Dam, \NPB{260}{1985}{295}. }
\refdef[poteramos]{ F.Figueirido and E.Ramos, \IJMPA{5}{1991}{1805}.}
\overfullrule=0pt
\unsectioned
\def\pubblock{ \line{\hfil\twelverm Preprint-QMW-PH-93-1}
               \line{\hfil\twelverm US-FT/1-91}
               \line{\hfil\twelverm January 1993}
               \line{\hfil hep-th/9303034}}
\titlepage
\title{DIFFEOMORPHISMS FROM HIGHER DIMENSIONAL $\W$-ALGEBRAS}
\authors
\hfil{\caps Fernando~Mart\'\i nez
Mor\'as$^1$\email{fernando@gaes.usc.es},
Javier~Mas$^1$\email{jamas@gaes.usc.es}, and
Eduardo~Ramos$^2$\email{ramos@v2.ph.qmw.uk}}\hfil
\addresses
$^1${\it Departamento de F{\'\i}sica de Part{\'\i}culas Elementales,
Facultad de F{\'\i}sica, Universidad de Santiago, Santiago de
Compostela 15706, SPAIN}\hfil\break\noindent
$^2${\it Department of Physics, Queen Mary and Westfield College,
Mile End Road, London E1 4NS, UK} %

\abstract{Classical $\W$-algebras in higher dimensions have been
recently constructed. In this letter we show that there is a
finitely generated subalgebra which is isomorphic to the
algebra of local diffeomorphisms in $D$ dimensions. Moreover, there is
a tower of infinitely many fields transforming under this subalgebra
as symmetric tensorial one-densities. We also unravel a structure
isomorphic to the Schouten symmetric bracket, providing a natural
generalization of $\w_\infty$ in higher dimensions.}

\endtitlepage

\subsection{Introduction}

The purpose of this letter is to give a simple account of
$D$-dimensional classical $\W$-algebras and their intimate connection
with the algebras of local diffeomorphisms of a $D$-dimensional manifold.

In general, classical one dimensional $\W$-algebras are defined as nonlinear
extensions of diff$(S^1)$ by tensors of integer weights. These algebras
appear naturally in the context of two dimensional conformal field theory. They
are obtained via the centerless $c\rightarrow\infty$ limit of the OPE's in
theories enjoying $\W$ symmetry. The canonical example of such a system is
provided by the 3-state Potts model and its $\W_3$-symmetry. The classical
$\w_3$ algebra associated with it is explicitely given by
$$\eqalign{\pb{T(x)}{T(y)}=&-(T\d +\d T)_x\cdot\delta (x-y)\cr
\pb{W(x)}{T(y)}=&-(2W\d + \d W)_x\cdot\delta (x-y)\cr
\pb{W(x)}{W(y)}=&~({2\over 3}T\d T)_x\cdot\delta (x-y)\cr}$$

It was shown in \[ClassLim] that these classical $\W$-algebras also appear as
Poisson structures in the commutative limit of the ring of pseudodifferential
operators in one dimension. It was precisely this relationship which
allowed two of the present authors to generalize this construction to higher
dimensions. Nevertheless, one crucial point was missing in \[Mamoramos].
Although conjectured, it was not proven that these algebras are extensions of
higher dimensional diffeomorphisms algebras. We will show in what follows that
this is indeed the case, and therefore that these new algebraic structures
fully deserve their name.

Before getting into more technical matters, we would like to point out
that these classical $\W$-algebras provide hamiltonian
structures for dispersionless KP-type hierarchies \[Mamoramos]. In one
dimension these hierarchies play a fundamental role in
the planar limit of  non-critical string theory with $c\leq 1$, as well as
in topological models. It is our hope that these new
structures will come into play in the higher dimensional descriptions of these
physical problems. We believe that the integrability of the associated
hierarchies as well as the relationship to diffeomorphism algebras, support
(though weakly) our expectations.

In what follows we have tried to avoid, as much as possible, to get into
too technical a description of the subject. We refer anyone
who wishes a detailed analysis of the general formalism to \[Mamoramos] and
references therein.

\subsection{The recipe}

The natural arena for the construction of
higher dimensional classical $\W$-algebras is
provided by a phase space $Y^{2D}$ with
coordinates $(x^i,\xi_i )$ with $i=1,\cdots ,D$,
and canonical Poisson bracket given by

$$\pb{f}{g} =\pder{f}{\xi_i}\pder{g}{x^i}
-\pder{f}{x^i}\pder{g}{\xi_i}.\(pebes)$$

{}From now on we will restrict ourselves to
homogeneous functions on $Y^{2D}$, where the degree of homogeneity
is defined as follows: a function is said to be of degree
$n$ if under the rescaling $\xi_i\rightarrow t\xi_i\ \forall\
i=1,\cdots ,D$
$$f(x^i,\xi_i)\rightarrow f(x^i,t\xi_i)=t^n
f(x^i,\xi_i).$$

Let us now define the symplectic trace \[Guill]\[Wod] as follows
$$\Tr f=\int dx^D d\Omega_{\xi} ~ f(x,\theta_\xi)$$
if $f$ is of degree $-D$ and zero otherwise.
$d\Omega_{\xi}$ stands for the standard measure of the
$D-1$ sphere in the $\xi$ coordinates. The notation
is justified because  of the ``trace" property\fnote{
We are assuming here that our $x$-space is compact or, equivalently,
that our type of functions decay fast enough at infinity.}
$$\Tr \pb{f}{g} =0.$$

With this machinery it is now simple to construct the analogs
of the classical $W$-algebras in arbitrary dimension.
Let us define the formal generating functional
$$\Lambda =\xi^m +\sum_{j=1}^{\infty} U_j(x^i,\xi_i),$$
where $m>0$, $\xi =(\sum_{k=i}^{D}\xi_k^2 )^{1\over 2},$
and the $U_j$'s are homogeneous functions of degree $m-j$.
Therefore $\Lambda$ can be rewriten as
$$\Lambda=\xi^m +\sum_{j=1}^{\infty} u_j(x^i,\theta_{\xi})
\xi^{m-j},$$
with $\theta_{\xi}$ denoting the angular coordinates associated
with the $\xi_i$.
The Poisson brackets among the $u$'s are defined via the
generalized classical Adler map $J$ (defined below) and linear
functionals on $\Lambda$. The latter are given by
$$
F_X=\Tr X\Lambda.\(pedo)
$$
It is obvious from this and the properties of the trace,
that the most general $X$ defining a nontrivial functional
is of the form
$$X=\sum_{j=1}^{\infty}X_j(x_i,\theta_{\xi})\xi^{j-m-D}.$$

We can now define the ``Gel'fand-Dickey" brackets by
$$
\pb{F_X}{F_Z}_{\GD} =\Tr J(X) Z, \(Geldi)
$$
with
$$
J(X)=-\pb{\Lambda}{X}_{\ominus}\Lambda +
\pb{\Lambda}{(\Lambda X)_-}, \(clasadler)
$$
where the subscripts stand for the following projections,
if $Q=\sum_{k\in\integ} q_k$ with $q_k$ a homogeneous functions
of degree $k$, then
$$
Q_-=\sum_{k\leq 0} q_k \;\;\;\;\;\;\;
\hbox{and} \;\;\;\;\;\;\; Q_{\ominus}=\sum_{k\leq -D-1}q_k.
$$
We also define $Q_+=Q-Q_-$ and $Q_{\oplus}=Q-Q_{\ominus}$.
Notice by the way that the + ($-$) projection is the dual,
with respect to the symplectic trace, of the $\ominus$
($\oplus$) projection. Moreover, the + and $-$ projections
are subalgebras with respect the canonical Poisson bracket
defined by \(pebes). This comes out because, as the
reader can easily check, if the functions $f$ and $g$
have degrees $p$ and $q$ respectively then $\pb{f}{g}$
has degree $p+q-1$. We would like to remark that our
$+$ splitting differs from the usual one because we are excluding
from it the components of zero degree. This is required in $D>1$
if we want to preserve the subalgebra property described above.

  Because of its definition, and the grading
properties of the the canonical Poisson bracket,
$J(X)$ is bound to have the following  form:
$$J(X)=\sum_{i,j=1}^{\infty} (J_{ij}\cdot
X_j)\xi^{m-i},\(formofj)$$
where $J_{ij}$ is a differential operator with
coefficients that are at most quadratic in the $u$'s and
their derivatives. This together with \(pedo) and \(Geldi) imply
$$ \pb{u_i(x^i,\theta_{\xi})}{u_j(y^i,\theta_{\xi}')}_{\GD}
=-J_{ij}\cdot {\delta}^D (x-y)\delta (\Omega -\Omega
'),\(bra)
$$ with $\delta (\Omega -\Omega ')$ the delta
function associated with the standard measure in $S^{D-1}$.

%It is clear that this brackets are antisymmetric, but
%it is far from obvious that they obey Jacobi identities;
%nevertheless it was proven in \[Mamoramos]
%that this is indeed the case.

Although far from obvious, it is a main result of \[Mamoramos]
that these brackets define full fledged Poisson brackets.

It is posible to deform the generalized classical Adler
map by $\Lambda\to \Lambda +\lambda$, with $\lambda$ an
arbitrary constant, and obtain two different  Poisson
structures. Explicitely
$$J(X)\rightarrow J^{(2)}(X)+\lambda J^{(1)}(X). \(defor)$$
where
$$
\eqalign{
J^{(2)}(X) &= J(X)   \cr
J^{(1)}(X)&=-\pb{\Lambda}{X}_{\ominus}
+\pb{\Lambda}{X_-}. \cr }  \(primsegun)
$$
The two Poisson stuctures induced by $J^{(2)}$ and $J^{(1)}$
are said to be coordinated since any linear combination of them
is still a Poisson bracket.
For
``perverse" historical reasons they are commonly known as the ``second"
and ``first Gel'fand-Dickey brackets" respectively. Notice that,
by construction, the first structure induces brackets which are linear in
the $u_j$'s.

The algebraic structures that we are interested in are only
going to appear after imposing certain constraints in the form
of the operator $\Lambda$. Explicitely, we are going to set
$u_1=\cdots =u_D=0$ for the second structure, and
$u_1=\cdots = u_{m+D}=0$, where $m$ is the leading order in $\Lambda$,
for the first structure.

The constraint on $J^{(2)}$ is second class, and its implementation
follows Dirac's prescription, which in this particular case reads
$$
J^{(2)}_{ij}\rightarrow J_{ij}-\sum_{m,n=1}^{D} J_{in}
J^{-1}_{nm} J_{mj}\quad\forall\   ij>D
$$
where $J^{-1}$ is the inverse of the $D\times D$ matrix with entries
given by the $J_{nm}$ in \(formofj) and $1\leq n,m\leq D$. It is
worth pointing out that in spite of potential non-localities, because
of the term in $J^{-1}$, the resulting Poisson brackets are local, as a
straigthforward computation shows.

The constraint on $J^{(1)}$ is much more easily implemented by noticing
that the Poisson brackets of the $u_j$'s with $j\geq m+D$ and the
constraints are zero weakly, i.e. after imposing the constraints.
 Therefore in this case
$$J^{(1)}\rightarrow  -\pb{\Lambda}{X_+}_{\ominus}.\(kredu)$$

Notice that after the reduction the linear part of $J^{(2)}$
is given by
$$J^{(2)}_{linear}=-\pb{\Lambda}{X}_{\ominus}\xi^m
+\pb{\xi^m}{(\Lambda X\xi^{-1})_{\ominus}\xi }, \(linearpart)$$
which can be seen to be isomorphic to \(kredu) under the map $\Lambda
\rightarrow \Lambda \xi^{-m}$. Another important fact
is that, upon the imposition of the
constraints in the second structure, the Poisson brackets
involving the field $u_{D+1}$ with any other of the $u_j$'s are
directly linear, therefore isomorphic to the ones obtained using the
first structure after the relabeling that maps $u_j\rightarrow
u_{j-m}$. This is a crucial feature which drastically simplifies
the explicit construction of the diffeomorphism subalgebra from the
second Gel'fand-Dickey brackets.

The $D=1$ and $D=2$ Poisson brackets defined by \(bra) were
explicitely computed in \[Mamoramos] and we now briefly
describe their main features.

For $D=1$ and $m=1$ the second structure\fnote{It can be shown using the
techniques developed in \[ClassLim] that for any $D$ the second structure
is isomorphic for all values of $m\neq 0$, while the first can be
shown to be independent of $m$ by explicit computation.}
coincides with a limit
$n\rightarrow\infty$ of the classical $\w_n$ algebras
after setting the constraint $u_1=0$. We would like to
stress that this limit is intrisically non-linear and
therefore non-isomorphic to the standard $\w_{\infty}$.
However, the first structure, after imposing the constraint
$u_1=u_2=0$ turns out to be exactly $\w_{\infty}$, as
expected.

For $D=2$ it was shown that there is a subalgebra
isomorphic to the algebra of diffeomorphisms in two dimensions.
This subalgebra
plays an analogue role to the one of
Virasoro in the one dimensional case. Moreover, it was shown that
the first structure has a subalgebra generated by  symmetric
tensor-one-densities that  offers a natural generalization
of $\w_\infty$ in two dimensions \[Hull2], and is related to the Schouten
bracket. In what follows we will show that this lower dimensional
properties extend for arbitrary dimension $D$.

\subsection{The algebra of diffeomorphisms}

Let us begin by recalling some simple facts about diffeomorphisms.
In local coordinates infinitesimal diffeomorphims are defined through
the map $x^\mu \to x^\mu + \epsilon f^\mu(x)$.  The algebra
generated by these transformations is isomorphic to the algebra of
vector fields $\vec f \equiv f^\mu(x) \d_\mu$, {\ie}
$$
\comm{\vec f}{\vec g} = (f^\mu(\d_\mu g^\nu) -g^\mu(\d_\mu f^\nu) )(x)
\d_\nu  .\(vectn) $$

In a field theory invariant under diffeomorphisms the above algebra
will be implemented via Poisson brackets\fnote
{Strictly speaking, in  classical field theory only spatial
diffeomorphisms will be represented via Poisson brackets in this way.
The splitting between spatial and time coordinates required in
the canonical formalism is obviously not invariant under arbitrary
diffeomorphisms and, moreover, requires the introduction of a metric.
Consequently the Poisson brackets among spatial and
timelike diffeomorphisms have, in general,
an  explicit dependence on the metric.
In this sense $D$ should be considered the dimension of the spacelike
coordinates. Nevertheless, this should not concern us for the abstract
manipulations that follow.}
, {\ie} there must be
a Lie algebra homomorphism given by

$$
\vec f=f^\mu(x) \d_\mu \to Q_{\vec f}= \int d^D x
f^\mu(x) P_\mu(x). \(mapvecfun)
$$
where the $P_\mu$ are the generators of infinitesimal diffeomorphisms
and such that
$$ \pb{Q_{\vec f}}{Q_{\vec g}}_{PB} =
                  Q_{\comm{\vec f}{\vec g}}.\(isomo) $$

For the left hand side of \(isomo) we get
$$
\pb{Q_{\vec f}}{Q_{\vec g}}_{PB}
= \int d^Dx \int d^Dy ~f^\mu(x) g^\nu(y)
{}~\pb{P_\mu(x)}{P_\nu(y)}_{PB}  \(lehand)
$$
whereas for the right hand side, because of \(vectn), we
should get
$$
Q_{\comm{\vec f}{\vec g}} = \int d^Dx (f^\mu(\d_\mu g^\nu)
-g^\mu(\d_\mu f^\nu) )(x) P_\nu(x). \(rihand)
$$
Equating both sides we obtain \fnote{ We are
only considering, as is usually the case, diffeomorphisms with
compact support, so boundary terms can be consistently neglected.}
 $$
\pb{P_\mu(x)}{P_\nu(y)}_{PB} = ( P_\nu \d_\mu + \d_\nu P_\mu)_x\cdot
\delta(x-y). \(algdif)
$$
Notice that for dimension one \(algdif) is, up to an irrelevant global sign,
nothing but the centerless Virasoro algebra.

Let us first show how to obtain \(algdif) from the first
Gel'fand -Dickey bracket. After the reduction of setting the first
$m+D$ fields equal to zero, the brackets are obtained from
\(Geldi) with  $J^{(1)}$
given in \(kredu).
To the vector field $\vec f$ we associate the linear functional
$Q_{\vec f}$
$$ \vec f=f^\mu(x) \d_\mu \to Q_{\vec f} = -\Tr ~f^\mu \xi_\mu~
\Lambda.\(key)$$
Comparing \(key) with \(mapvecfun) we obtain
$$
P_{\mu}(x)= ~- \int d\Omega_{\xi} ~\xi_{\mu}
u_{D+m+1}(x,\theta_\xi) .\(pemu)
$$
 Moreover from \(pedo) it follows that $Q_{\vec f} = F_{X_{f}}$
with $X_{f}=-f^{\mu}\xi_{\mu}$. Therefore
$$ \eqalign{\pb{Q_{\vec f}}{Q_{\vec g}}^{(1)}_{\GD}
&=\Tr J^{(1)} (X_f)X_g \cr
&= \Tr \pb{f^\mu\xi_\mu}{\Lambda}_\om ~g^\nu(x)\xi_\nu \cr
&= \Tr \pb{f^\mu\xi_\mu}{\Lambda} ~g^\nu(x)\xi_\nu \cr
&=-\Tr \pb{f^\mu\xi_\mu}{g^\nu(x)\xi_\nu} ~\Lambda \cr
&= -\Tr (f^\mu(\d_\mu g^\nu) - g^\mu(\d_\mu f^\nu))~ \Lambda \cr
&= Q_{\comm{\vec f}{\vec g}}. \cr }\(sifu)
$$
This implies, as before, that the first Gel'fand-Dickey brackets
of the $P_{\mu}$ defined by \(pemu) reproduce the algebra of
generators of infinitesimal diffeomorphisms in $D$ dimensions.

Furthermore, we can extend the
map defined by \(key) to symmetric contravariant tensors of higher rank.
Under diffeomorphisms, a rank $r$ contravariant symmetric tensor transforms
infinitesimally with the Lie derivative  $T^{\mu_1,...,\mu_r} \to
T^{\mu_1...\mu_r} + \epsilon (\lied{\vec f} T)^{\mu_1...\mu_r}$, where
$$
(\lied{\vec f} T)^{\mu_1...\mu_r} =
f^\nu \,( \d_\nu  T^{\mu_1...\mu_r}) - r
T^{\nu,(\mu_1...\mu_{r-1}}\, \d_{\nu} f^{\mu_r)},\()$$
and the brackets on the superindices stand for symmetrization.

Let us therefore define the associated functional in the form:
$$
T \to Q_T  =- \Tr ~T^{\mu_1...\mu_r}\xi_{\mu_1}...\xi_{\mu_r}\,
 \Lambda. \(homlinten)
$$

An easy computation parallel to the one in \(sifu)
yields
$$\pb{Q_{\vec f}}{Q_T}^{(1)}_{\GD} =  	Q_{{\cal L}_{\vec f} T}.\(vecfu)$$

Moreover, if we define $Q_{R}$ via the components of a contravariant
tensor of order $s$,
$$\pb{Q_R}{Q_T}^{(1)}_{\GD} =Q_{\comm{R}{T}_{\Sch}},\(chota)$$
where $[R,T]_{\Sch}$ stand for the symmetric Schouten bracket
\[Nijenhuis], which in
a coordinate basis reads
$$[R,T]_{\Sch}^{\mu_1...\mu_{r+s-1}}=
s R^{\nu,(\mu_1..\mu_{s-1}}\d_\nu
T^{\mu_s..\mu_{r+s-1})}-
r T^{\nu,(\mu_1..\mu_{r-1}}\d_{\nu}
R^{\mu_r..\mu_{r+s-1})}$$

This implies, as
before, that if we define
$$
P_{\mu_1...\mu_r}=-\int d\Omega_{\xi}~\xi_{\mu_1}\cdots
\xi_{\mu_r} u_{D+m+r}\(pemunu).
$$
the first Gel'fand-Dickey bracket among the $P$'s
define a closed subalgebra given by
$$\eqalign{
\pb{P_{\mu_1...\mu_r}(x)}{P_{\mu_{r+1}...\mu_{r+s}}(y)}_{\GD}^{(1)}=
 &~ \bigl( \, \sum_{j=1}^r P_{\mu_1..\hat\mu_j..\mu_r..\mu_{r+s}}
\d_{\mu_j}+   \cr
\sum_{j=1}^s \d_{\mu_{r+j}} &  P_{\mu_1..\mu_r..,\hat\mu_{r+j},..,
\mu_{r+s}} \bigr)_x \cdot\delta (x-y),\cr
}  \(chori)
$$
where the subindex with a hat is omitted. In particular,  the above
equation implies that $P_{\mu_1...\mu_s}$ transforms under diffeomorphisms
as a $s$-covariant symmetric tensorial one-density.

The analysis of the second Poisson structure simplifies considerably
if we make use of the isomorphism $u_j\rightarrow u_{j-m}$ between the linear
part of $J^{(2)}$ and $J^{(1)}$ mentioned
in \(linearpart).
 One can see with little effort that the second Gel'fand-Dickey bracket
involving the first nonzero field, $u_{D+1}$, and any other higher field
$u_{i>D+1}$ is linear. Henceforth,  we already know
the expression for all these brackets invoking the above mentioned isomorphism:
$$\pb{u_{D+m+1}}{u_{D+m+k}}^{(1)}_{\GD}\ \rightarrow
\ \pb{u_{D+1}}{u_{D+k}}^{(2)}_{\GD}.$$
Therefore, the key properties \(sifu) and \(vecfu) still hold for the
second Gel'fand-Dickey brackets, whereas for \(chota) this is not the case due
to the quadratic terms involved.

Summarizing, we have unraveled a set of similarities among the higher
dimensional classical $\W$-algebras constructed in \[Mamoramos] and the
standard one-dimensional algebras of the $\w_{\infty}$ type. We have
shown explicitly how to construct a finitely generated subalgebra
isomorphic to the algebra of diffeomorphisms in $D$-dimensions, therefore
these new structures can be naturally understood as extensions of the
symmetry algebra for generally covariant theories. Moreover, we have also
shown that there is an infinite tower of fields $P_{\mu_1...\mu_k}$
transforming as $k$-covariant one-densities.
Nevertheless, as expected, there are some relevant differences between
the one and higher dimensional case. The most important
one is that, as was explicitly shown in \[Mamoramos], there are also
fields transforming as infinite dimensional reps. of the diffeomorphism
subalgebra. Here, we should clearly distinguish between the first and
second Gel'fand-Dickey brackets. Whilst in the first the
$P_{\mu_1\cdots\mu_k}$ form a closed subalgebra which naturally generalize
$\w_{\infty}$ to higher dimensions \[Hull2], in the second bracket
this is not the case. It is not clear to us at this point if these extra
fields are intrinsically necessary for the closure of the nonlinear algebra,
or if they can be wiped out by some suitable hamiltonian reduction.
Another important difference lies in the difficulty in constructing
an analog of $\w_n$ in higher dimensions. This should be reminiscent
of the work in \[FradBerends], where it was shown that one way to construct
consistent theories involving higher spin gauge fields is through the
introduction of an infinite number of fields with all possible spins.
 However, this remark must be taken with due care, since
our algebras represent Hamiltonian structures, and therefore should rather be
related to the spatial part of diffeomorphism invariant field theories, such as
for example canonical gravity. Yet the fact that these infinite
dimensional algebras arise as hamiltonian structures of integrable systems,
points in the direction of an algebraical, or conformal-field-theory-like
approach, in the  search for the quantization of those theories.
\bigskip
{\bf Acknowledgments}

We are grateful to J. M. Figueroa O'Farrill, J. Petersen
and C. M. Hull for a
careful reading of the manuscript. ER takes great pleasure in
expressing his thanks to the Physics Dept. of the University
of Santiago for its hospitality.

\refsout

\bye